\documentstyle[12pt,aasms]{article}
\newcommand{\lae}{\mathrel{<\kern-1.0em\lower0.9ex\hbox{$\sim$}}}
\newcommand{\gae}{\mathrel{>\kern-1.0em\lower0.9ex\hbox{$\sim$}}}

\newcommand{\Msun}{\mbox{ M}_{\odot}}
\newcommand{\Ni}{${}^{56}\mbox{Ni}$}
\newcommand{\Fe}{${}^{56}\mbox{Fe}$}
\newcommand{\Co}{${}^{56}\mbox{Co}$}

\newcommand{\beq}{\begin{equation}}
\newcommand{\eeq}{\end{equation}}

\begin{document}

\title{A New Radiation Hydrodynamics Code and Application to the
Calculation of Type Ia Supernovae Light Curves and Continuum Spectra}
\author{Xiao-he Zhang and Peter Sutherland}
\affil{Department of Physics \& Astronomy\\ McMaster University\\
Hamilton, Ontario, Canada}

\begin{abstract}
A new, fully dynamic and self-consistent radiation hydrodynamics code,
suitable for the calculation of supernovae light curves and continuum
spectra, is described. It is a multigroup (frequency-dependent) code
and includes all important $O(v/c)$ effects. It is applied to the
model W7 of Nomoto, Thielemann, \& Yokoi (1984) for supernovae of
type Ia. Radioactive energy deposition is incorporated through
use of tables based upon Monte Carlo results. Effects of line
opacity (both static or line blanketing and expansion or line
blocking) are neglected, although these may prove to be important. At
maximum light, models based upon different treatments of the opacity
lead to values for $M_{B,max}$ in the range of -19.0 to -19.4. This range
falls between the values for observed supernova claimed by Leibundgut
\& Tammann (1990) and by Pierce, Ressler, \& Shure (1992).
\end{abstract}


\clearpage
\section{Introduction}

Until very recently, no unified approach to the modeling of
supernovae light curves and spectra has been pursued. For stars (even
those with steady winds) sophisticated treatments of the atmosphere
and line formation are by now routine. In the case of supernovae there
is dynamical coupling of the radiation and the matter (at maximum
light $t_{diffusion} \sim t_{dynamic} \sim R/v \sim 10^6 \mbox{ s}$),
much of the ejecta are semi-transparent (and the radiation and matter
are not in thermal equilibrium), there are non-trivial $O(v/c)$
effects, and there are uncertainties about the opacity. Add to this
uncertainties about the specificity of the model of the exploded star
and restrictive limitations on computer resources, and it is
unsurprising that few attempts have been made to develop and use a
fully-integrated and self-consistent radiation hydrodynamics code (but
see Falk \& Arnett 1977). Instead, depending on the goal of the study,
the radiation dynamics has been reduced to either gas dynamics with
radiation diffusion or radiative transfer in the approximation of a
quasi-steady wind. The former approach (see, for example, Sutherland
\& Wheeler 1984) has been used to calculate light curves for assumed
models of the explosion, for which the key issues are rise time,
maximum brightness and color, and rate of decline from the peak. In
such calculations the opacity is often taken to be gray, uniform, and
constant (typically, $\kappa \sim 0.1 \mbox { cm}^2\mbox{ g}^{-1}$
independent of density, temperature, composition, and frequency). The
latter approach (Branch 1990, Harkness 1986, Wheeler \& Harkness 1990)
has been used to generate realistic synthetic line spectra that can
serve as diagnostics for the velocity and density profiles of the
supernova. For both the work of Branch, in which the Sobolev or
escape-probability approximation is adopted in calculating P Cygni
profiles, and the work of Harkness, which solves the relativistic,
comoving frame radiative transfer equations as a function of
frequency, an inner boundary condition is required. This must
either be guided by observations or be given by some other, dynamical,
calculation.

Recently Ensman (1991) (see also Ensman \& Burrows 1992),
H\"{o}flich, Khokhlov, \& M\"{u}ller (1991) (see also
Khokhlov, M\"{u}ller, \& H\"{o}flich 1992,1993; H\"{o}flich, M\"{u}ller,
\& Khokhlov 1993) have presented calculations based upon new
radiation hydrodynamics codes. In both cases the dynamical equations
are solved in the comoving frame and include all $O(v/c)$ effects.
Ensman works with the frequency-integrated moment equations and
Rosseland mean opacities. The Eddington factor (which relates the
second radiation moment to the zeroth moment and is essential for
closure) is obtained by solving a {\em steady} model transfer
equation, which also provides the surface boundary condition for the
flux. In effect this code is a ``two-temperature'' code, one each for
the matter and radiation, and is suited to light curve calculations
but not to the calculation of colors and continuum spectra. The code
of H\"{o}flich et al. (1991) deals with the full frequency-dependent
equations, and thus yields both light curves and continuum spectra. It
also incorporates the expansion opacity (Karp et al. 1977). The
expansion of the matter is assumed to be simply homologous;
consequently no dynamical interaction of the radiation and gas is
allowed. This is probably an adequate approximation for Type Ia
supernovae (SNIa's) but may be vitiated for Type II supernovae
(SNII's) models based upon extended red supergiant progenitors.

In this paper, we present a one-dimensional, comoving frame,
Lagrangian radiation hydrodynamics code that will (1) solve a
monochromatic model radiative transfer equation (correct to $O(v/c)$
to obtain the frequency-dependent Eddington and sphericity factors and
the flux surface boundary condition; (2) solve the
frequency-integrated radiation moments equation -- correct to $O(v/c)$
-- for radiation energy density and flux; (3) solve the system of
monochromatic radiation moments equations to form the radiation energy
and flux spectra; (4) use the computed radiation and flux spectra and
precalculated, composition sensitive opacity tables, averaged in a
series of frequency bins, to form time-dependent frequency means of
opacities; and (5) coupled with an equation of state for ideal
electron/ion gas and a self-consistent solution of the Saha equation
for ion balance, solve the radiation hydrodynamics equation for gas
energy and radial velocity.  This represents an improvement over both
the work of Ensman and H\"{o}flich et al. in that the treatment is
both fully dynamic and frequency dependent. In addition, the frequency
means of the opacity are more self-consistent because the computed
radiation energy and flux spectra were used in these means.  In
\S II, we summarize the formalism employed in our light curve
calculations and the important physics inputs to the radiation
hydrodynamics code.

There are several reasons to calculate the light curves for models of
SNIa's using the best radiative dynamics code and the best input
physics. Observationally, the light curves of SNIa's appear to be
remarkably homogeneous (see the atlas of Leibundgut et al. 1991b) and
are strong candidates as ``standard candles.'' There continues to be
some dispute over the normalization of these objects. On the one hand,
Leibundgut \& Tammann (1990) find for SNIa's in the Virgo cluster that
at B maximum light $M_B = -19.8 \pm 0.12$. (Jacoby et al. 1992 survey
the calibration of SNIa's and reports similar results for other
methods and samples.) On the other hand, a new determination of the
distance to IC 4182 based upon I-band and K-band photometry of the
brightest red supergiants implies $M_{B,max} = -18.8 \pm 0.34$ for SN
1937C. These two contrasting results correspond to approximate values
for $H_0$ of 50 and 85 $\mbox{km}\mbox{ s}^{-1}\mbox{ Mpc}^{-1}$. Thus
light curve calculations (and by this is meant, at the least, light
curves for each of the principle bands, U, B, V, \ldots) have the
potential to constrain, perhaps reject, either the assumed model for
the SN or a range of values for $H_0$ if SNIa's have uniform
properties. The other major reason for calculating SNIa light curves
with modern codes is that the paradigm for SNIa's is well entrenched
and well defined. The prevailing paradigm is the explosion of a
carbon-oxygen white dwarf near the Chandrasekhar limit. For this model
it is essential that $\sim 1.0 \Msun$ of C/O be partially or fully
incinerated in order to explain the velocity profile implied by the
observed spectra (see, for example, Branch et al. 1982 for the case of
SN 1981B). Furthermore, $\gae 0.1 \Msun$ must be only partially
incinerated to intermediate mass elements. What remains unresolved in
the model is the exact amount of C/O fully incinerated to \Ni. Is the
material consumed by a detonation (Arnett 1969), a deflagration
(Nomoto, Sugimoto, \& Neo 1976) or a deflagration that turns into a
detonation (Woosley 1990, Khoklov 1991)? The total mass of \Ni,
through energy input by the trapping of $\gamma-$rays released in its
decay to \Co\ and thence to \Fe, is responsible for the
``normalization'' of the light curve; the rise time and width of the
peak reflect the kinetic energy of the ejecta which is already
constrained by the observed velocities. Light curve calculations based
upon overly simple approximations to the opacity (see, for example,
Sutherland \& Wheeler 1984) lead to the conclusion that $M_{Ni} \sim
0.8 \Msun$ if $M_{B,max} \sim -19.8$. This is to some extent confirmed
by the semiempirical claim advanced by Arnett, Branch, \& Wheeler
(1985) that, at maximum light, the bolometric luminosity equals the
{\em instantaneous} rate of energy deposition by \Ni\ decay. These
conclusions need to be tested by more rigorous calculations, which may
in turn lead to tighter constraints on the nature of the deflagration
and/or detonation and the amount of \Ni\ produced.

In \S III, we describe briefly the initial hydrodynamic model
used in our calculation and present the results of a few test runs
using this model for both ``gray'' and full frequency-dependent
computations.

\section{Description of the code}

The code described here is a one-dimensional (spherically symmetric),
Lagrangian one. All equations are expressed and solved in the comoving
frame (actually a series of frames, each instantaneously at rest with
respect to the local matter).  The radiation hydrodynamics equations
for the matter and radiation, and the radiative transfer equations,
are solved simultaneously following the prescription of Mihalas \&
Mihalas (1984).  All terms important to $O(v/c)$, where $v$ is the
radial velocity and $c$ is the speed of light, are retained.  The
emergent flux and its spectrum are transformed to the frame of the
observer only at the very last stage.

\subsection{Radiation Hydrodynamics}

In solving radiation hydrodynamic equations for SN light curves, a
large range of optical depth must be covered: from a mostly optically
thick envelope shortly after the explosion to almost transparent
ejecta at late times.  Near maximum light, the photosphere has already
receded significantly into the ejecta. It is not correct to assume at
this point that the radiation is still in thermal equilibrium with the
gas, a condition when the diffusion approximation applies.  In order
to faithfully follow the dynamic evolution of both gas and the
radiation field, it is imperative to treat them individually, without
any assumption about their thermal equilibrium.  The equations for the
evolution of the momentum and energy of the matter are

\beq
{{Dv}\over{Dt}}=-  \left({1 \over \rho} \right)
    {{\partial (p_g + Q) }\over{\partial r}}
 + \left({\chi_F\over c} \right) F_r - {{G m(r)} \over {r^2}}
\label{a}
\eeq

\beq
{{De} \over {Dt}} + (p_g +Q) {D \over {Dt}}\left( {1 \over \rho}\right) =
    c \kappa_E E_r - 4 \pi \kappa_P B(T)  + \epsilon
\label{b}
\eeq
where $D/Dt$ is the Lagrangian time derivative, $\rho$, $e$, $p_g$ and
$T$ are the density, energy per gram, thermal pressure and temperature
in the ejecta, $m(r) \equiv
\int 4 \pi r^2\rho(r) dr$ is the total mass within the sphere of
radius $r$, and $\epsilon$ is the heat source due to radioactive
decay. The gas absorbs momentum and energy from the radiation at the
rates $(\chi_F F_r /c)$ and $c\kappa_E E_r$, respectively, and
radiates energy at the rate $ 4\pi \kappa_P B(T)$. $B(T) \equiv (a
c/4\pi)T^4 $ is the Planck function at the matter temperature, $F_r$
and $E_r$ are the radiation flux and energy density, $\kappa_E$ is the
energy mean of the absorptive opacity $\kappa_\nu$, defined in
equation (\ref{n}) later in this section, $\kappa_P$ is the Planck
mean of the absorptive opacity, and $\chi_F$ is the flux mean of the
total opacity $\chi_\nu$, defined in equation (\ref{o}) also later in
this section.  Our one departure from the conventions of Mihalas \&
Mihalas (1984) is that all opacities have the units $\mbox{
cm}^2\mbox{ g}^{-1}$ rather than simply $\mbox{ cm}^{-1}$; that is,
our opacities need to be multiplied by $\rho$ to equal theirs.  To
handle shocks due to the initial model, an artificial viscosity term
for zone $k+1/2$ and at time $n$,

\beq
Q_{k+1/2}^n =\left\{
\begin{array}{ll}
 2 a^2 (v_{k+1}^n - v_k^n )^2 \rho_{k+1/2}^n
&\mbox{ for ${D \over {Dt}}\left( {1 \over \rho}\right) < 0 $,
 $ {{dv}\over{dr}} <0$ }\\
0 & \mbox{otherwise.}
\end{array}
\right .
\label{c}
\eeq
is also included.  Here $a$ is an adjustable parameter, generally set at
a value slightly larger than 1.

The evolution, in the comoving frame, of the frequency integrated
radiation moments is given by

\beq
 {{D} \over {Dt}} \left( {E_r \over \rho} \right)
 + \left[ f {D \over {Dt}}( {1 \over \rho})
   - (3 f -1 ) { v \over {\rho r }} \right] E_r
 =  4 \pi \kappa_P B - c \kappa_E E_r
    - {{\partial (4 \pi r^2 F_r)} \over {\partial m}}
\label{p}
\eeq

\beq
 {1 \over c^2} {{D F_r} \over {Dt}}
 + {1 \over q} {{\partial (f q E_r)} \over {\partial r}}
  =  - {{\chi_F \rho } \over c } F_r
     - {2\over c^2} \left( {v \over r} + { {\partial v} \over {\partial r}}
      \right) F_r
\label{q}
\eeq

These equations, and those that follow, are essentially identical to
those presented in \S 95 and \S 98 of Mihalas \& Mihalas (1984); see
their equations 95.18, 95.19, 98.1 and 98.2. There is one reduction
that we do not do: the term in each flux moment equation with
coefficient $2\left( {v \over r} + { {\partial v} \over
{\partial r}} \right) /c^2$ is {\em retained} for the following reasons.
This term is normally dropped because it is $O(v/c)$ relative to the
other terms (except for the $D/Dt$ term which is also of this order).
However, this coefficient is essentially just $4/(tc^2)$ because the
expansion is almost perfectly homologous, and near and definitely
after the light curve peak the $DF/Dt$ term is also $O(1/t)$ because
the radiation field is changing on a dynamical timescale.

In addition to the various opacities which require evaluation from
knowledge of (or assumptions about) the energy and flux spectral
profiles, there are two other elements that must be specified: the
Eddington factor $f$ (and the related sphericity factor $q$) and a surface
boundary condition on the relationship of the flux to the radiation
energy density.  [The boundary condition at $r=0$ is just $F_r = 0$.]
These factors (at frequency $\nu$) are defined by:
\beq
f_\nu \equiv {{\int I_\nu \mu^2 d \mu } \over {\int I_\nu d \mu}}
\label{d}
\eeq

\beq
\ln (q_\nu)=\int_{r_c}^r [(3f_\nu -1)/(r'f_\nu)]dr'
\eeq
where $I_\nu $ is the monochromatic radiation intensity at frequency
$\nu$, and $\mu \equiv \cos \theta$ the direction cosine of light rays
with respect to the outward radial direction.  The ``core'' radius
$r_c$ that enters the equation for $q_\nu$ is simply that radius
interior to which the matter is sufficiently optically thick so that
$f_\nu = 1/3$.  These factors are to be obtained as functions of
frequency, in view of the strong frequency-dependence of the opacity.
To obtain these essentially geometrical factors, the approach adopted
is to solve the frequency-dependent version of the model radiation
transfer equation proposed by Mihalas \& Mihalas (1984).  This model
transfer equation, which neglects ray curvature and Doppler shifts,
can be solved along parallel tangent rays through zone centers. It
decomposes into two equations for the symmetric and anti-symmetric
combinations:

\begin{eqnarray}
j_\nu (\mu) & \equiv & [I_\nu (r,\mu) + I_\nu (r, - \mu) ]/2
       \;\;\;\;\; 0\le \mu \le 1 \label{g}\\
h_\nu (\mu) & \equiv & [I_\nu (r,\mu) - I_\nu (r, - \mu) ]/2
       \;\;\;\;\; 0 \le \mu \le 1
\label{h}
\end{eqnarray}
yielding:
\begin{eqnarray}
 {1 \over c} {{Dj_\nu} \over {Dt}} + {{\partial h_\nu}\over {\partial s}}
& = & \kappa_\nu \rho B_\nu(T)
    - \left[ \kappa_\nu \rho + (1 - 3 \mu^2) { v \over {r c }}
    - {{1+\mu^2} \over c} {{D \ln \rho} \over {Dt}} \right] j_\nu \label{e}\\
 {1 \over c} {{D h_\nu} \over {Dt}} + {{\partial j_\nu}\over {\partial s}}
& = & - \chi_\nu \rho h_\nu     - {2\over c}
\left( {v \over r} + { {\partial v} \over {\partial r}} \right) h_\nu
\label{f}
\end{eqnarray}
Here $s$ is the distance along the ray from the front-back symmetry
plane ($s^2=r^2-p^2$ where $p$ is the impact parameter for a given
ray).  The boundary conditions used for these equations are that
$h_\nu$ is zero at the origin and at $\theta=\pi/2$ (the symmetry plane),
and that at the outer boundary ($r=R$) $h_\nu = j_\nu$ (no incident
radiation at the surface).  Then the frequency-dependent Eddington
factor $f_\nu$ is:

\beq
f_\nu = \int_0 ^1 j_\nu \mu^2 d \mu / \int_0 ^1 j_\nu d \mu
\label{i}
\eeq
and the ratio at the surface of flux to energy density, at frequency
$\nu$ is:
\beq
	F_\nu/(cE_\nu) = \int_0 ^1 h_\nu(\mu) \mu d\mu \left/
	\int_0^1 j_\nu(\mu) d\mu \right.
	= \int_0 ^1 j_\nu(\mu) \mu d\mu \left/
	\int_0 ^1 j_\nu(\mu) d\mu \right.
\label{jj}
\eeq
because of the surface boundary condition.

The opacities that enter equations (\ref{a}) and (\ref{b}) require various
spectral moments of the frequency-dependent opacity $\kappa_\nu$. The
model transfer equations above are suited to the calculation of
geometrical factors but are oversimplified for the calculation of the
spectra because of the neglect of a critical frequency derivative
(see equation [\ref{l}] below).  To estimate the radiation energy and flux
spectral profiles

\beq
e_\nu \equiv E_\nu / E_r
\label{j}
\eeq

\beq
\phi_\nu \equiv F_\nu / F_r
\label{k}
\eeq
it is necessary to solve the monochromatic radiation moments equation,
\begin{eqnarray}
{{D} \over {Dt}} \left( {E_\nu \over \rho} \right)
 + \left[ f_\nu {D \over {Dt}}( {1 \over \rho})
   - (3 f_\nu -1 ) { v \over {\rho r }} \right] E_\nu
& & \nonumber \\
 - {\partial \over {\partial \nu}} \left\{
      \left[ f_\nu {D \over {Dt}}( {1 \over \rho})
   - (3 f_\nu -1 ) { v \over {\rho r }} \right] \nu E_\nu\right\}
& = & \kappa_\nu ( 4 \pi B_\nu - c E_\nu )
    - {{\partial (4 \pi r^2 F_\nu)} \over {\partial m}}\label{l}
\end{eqnarray}
\beq
 {1 \over c^2} {{D F_\nu} \over {Dt}}
 + {1 \over q_\nu} {{\partial (f_\nu q_\nu E_\nu)} \over {\partial r}}
 =  - {{\chi_\nu \rho } \over c } F_\nu
  - {2\over c^2} \left( {v \over r} + { {\partial v} \over {\partial r}}
\right) F_\nu
\label{m}
\eeq
The boundary conditions for these equations are that $F_\nu = 0$ at
the origin and $F_\nu /( c E_\nu)$ at the surface is given by equation
(\ref{jj}).  For the frequency derivative term, the ``upwind'' scheme
in frequency space is used for stability.  Once the spectrum is found,
we form the energy means and the flux means of the opacity as in the
following,
\begin{eqnarray}
\kappa_E & \equiv & \int \kappa_\nu e_\nu d \nu \mbox{ ,}\label{n} \\
\chi_F & \equiv & \int \chi_\nu \phi_\nu d \nu \mbox{ .} \label{o}
\end{eqnarray}

In principle, one could use equations (\ref{a}), (\ref{b}), (\ref{l}),
and (\ref{m}) as the main system of dynamic equations and solve them
together for the hydrodynamic, thermal and radiation evolution.
Because all the important physics are included in equations (\ref{l})
and (\ref{m}), the frequency-integrated, or the bolometric, radiation
energy density and flux can be readily found from $E_\nu$ and $F_\nu$
by integrating them over frequency.  However, this will increase the
computational cost tremendously.  For economic reasons, we follow the
suggestions in Mihalas \& Mihalas (1984) and solve the ``gray''
(frequency-integrated) radiation moments equations instead. The main
system of dynamic equations will be (\ref{a}), (\ref{b}),
(\ref{p}),and (\ref{q}).  At each time step, we solve them implicitly
and iterate to convergence to determine the thermal, hydrodynamic, and
radiation structures inside the ejecta as accurately as possible.
This information is then used to solve equations (\ref{e}),(\ref{f}),
(\ref{g}),and (\ref{h}) implicitly for $j_\nu$, $h_\nu$, $E_\nu$ and
$F_\nu$.  From $E_\nu$ and $F_\nu$, estimates for $e_\nu$ and
$\phi_\nu$ are obtained.  Although equations (\ref{p}) and (\ref{q})
are the frequency-integrated form of equations (\ref{l}) and (\ref{m})
and the solution of equations (\ref{p}) and (\ref{q}) should be
equivalent to the solutions of equations (\ref{l})and (\ref{m})
integrated over frequency, it is inevitable that numerical errors in
these two forms of the radiation moments will accumulate at different
speed and eventually, the difference between $\int E_\nu d \nu$ and
$E_r$, for example, will be too big for equations (\ref{l}) and
(\ref{m}) to be consistent with equations (\ref{p}) and (\ref{q}).
Because it is more economical to implement more strict error control
in solving equations (\ref{p}) and (\ref{q}) than in equations
(\ref{l}) and (\ref{m}), we treat solutions of equations (\ref{p}) and
(\ref{q}) to be the ``true'' solution and treat any departure of $\int
E_\nu d\nu$ from $E_r$ as an accumulated error on the part of $E_\nu$.
When that happens, we rescale $E_\nu$ or $F_\nu$ according to $E_r$ or
$F_r$, keeping the spectral profiles unchanged.

\subsection{ Microphysics}

The microphysics components incorporated in the code are (1) the
equation of state (EOS), including ionization equilibrium, for the
matter, (2) the frequency-dependent opacities, and (3) the rate of
energy input from radioactive decay.

For the matter, the ideal gas EOS is valid shortly after the explosion
and certainly at all times when radiation transport is of interest.
(At early times, during and immediately after the explosion, the EOS
must allow for relativistic and partial degeneracy, etc.; for this
stage a modified version of the code of Wheeler \& Hansen (1971) was
used. Radiation transport then is insignificant or can be handled in
the diffusion approximation.) Ionization equilibrium is obtained from
the Saha equation, although the partition function was simplified to
include only ground state contributions. For a given composition, the
free electron fraction and ionization energy are precomputed and
tabulated for the density and temperature ranges appropriate to the
light curve calculation.

The frequency-dependent opacities were provided by Los Alamos National
Laboratory (LANL) (see Huebner et al. 1977; Magee, Merts, \& Huebner
1984). For a given composition, the multigroup opacities are tabulated
on a grid of 39 frequencies (0.5 eV -- 200 eV) by 11 temperatures (1
-- 10 eV) by 25 densities ($1.15\times 10^{-13}$ -- $1\times
10^{8}\mbox{ g cm}^{-3}$) although the temperature-density plane is
not fully sampled. Separate tables were provided for scattering and
absorption. The LANL opacities are not ideal for supernovae
calculations because they do not extend to sufficiently low densities
or temperatures. Opacity calculations that incorporate better physics
and have a wider domain of application have been implemented (Rogers
\& Iglesias 1992) but tables for appropriate compositions have not yet
been made available. The major shortcoming in the treatment of
opacities in the present calculations is the neglect of lines. The
large number of lines in the UV will be responsible for both line
blanketing (an essentially ``static'' effect dependent upon the high
density of lines and the overlap of successive lines by their natural
and/or thermal widths) and the expansion opacity (Karp et al. 1977),
also called line blocking. Harkness (1991) has successfully modeled
the spectra for SN 1981B near maximum light with the model W7, using
an approximate treatment of only line blanketing. The expansion
opacity is an effective continuum opacity due to scattering by a large
number of lines in a medium with a large velocity gradient and may
play a significant role in the UV.  The expansion opacity,
even as calculated by Karp et al. (1977) by treating the expansion of the
medium (relative to a chosen point) as infinite, homogeneous, and
isotropic (``Hubble flow''), places extraordinary demands on computing
resources. For the composition of SNIa's, $\gae 10^6$ lines need to be
retained, the ionization and excitation of many species and levels
must be obtained (for each zone in the model, and at different times
as conditions change significantly), and the effects of all lines must
be summed for each value of the expansion parameter ($s \propto
\rho/(dv/dr)$). The necessary computer resources were not available to
pursue these calculations. However, the basic assumption of Karp et
al. in treating SNs ejecta in the ``Hubble flow'' approximation
is limited. A local continuum opacity results from the sum of
probabilities for scattering in lines. A nominal scattering in a
strong line that contributes to the local expansion opacity at a
certain frequency, however, may take place at a remote point because
of the redshift required for the scattering atom.  Then the
assumptions of isotropy and homogeneity in the flow may be
unwarranted. In short, when expansion opacity effects are important,
they may not be simply calculable in terms of a local opacity.

The final microphysics ingredient in the code is the rate of energy
deposition due to radioactive decay. In the decay of \Ni\ and \Co, the
$\gamma$-rays released by the daughter nuclei Compton scatter (perhaps
repeatedly) off free and bound electrons, and less commonly produce
electron-positron pairs which are presumed to annihilate locally. The
result of the decays is thus a number of high-energy ($\lae 1$ MeV)
electrons. In all light curve calculations, except possibly at late
times when the ejecta are nearly transparent, it is a reasonable
assumption that these energetic electrons thermalize locally since the
electron mean free path is considerably shorter than that of the
originating $\gamma$-ray. What remains to be determined, then, is the
distribution throughout the ejecta of the energy deposited by the
Compton scattering of the $\gamma$-rays. Two approaches to the
calculation of the so-called deposition function have been employed in
the past: (1) a pure absorption model for $\gamma$-ray transport, as
advocated by Sutherland \& Wheeler (1984), and (2) a detailed, and in
principle exact, calculation through use of Monte Carlo methods (see
Ambwani \& Sutherland 1988). The latter requires many CPU cycles to
achieve a reasonable level of accuracy. The former can be dealt with
very efficiently by using a simple model gray atmosphere code. (The
source function is just the local mass fraction of radioactive
material and its distribution can be arbitrary.) We have compared the
results of the two approaches (for the model gray atmosphere
calculation we used the code of Swartz, Harkness, \& Wheeler 1991) and
found that at no epoch of interest and nowhere within the ejecta did
the deposition functions differ by more than 5\%, and the global means
were even closer. In view of the other uncertainties in modeling
SNs, it is clear that the deposition function can be calculated
with sufficient accuracy with the model gray atmosphere approach; the
Monte Carlo approach is overkill.

\section{Calculation of Model Light Curves and Continuum Spectra}

The SNIa model chosen for the calculation of light curves and spectra
is a somewhat mixed version of model W7 of Nomoto, Thielemann, \&
Yokoi (1984). This is a carbon deflagration model that yields an
acceptable light curve (as calculated in the diffusion approximation
with a uniform, gray, opacity) and for which synthetic spectra based
on P Cygni lines calculated for the velocity and composition profiles
of W7 (the latter requires some mixing) compares favorably with those
observed for SN 1972E and SN 1981B near maximum light (Branch et al.
1985). W7 is an exploded white dwarf of total mass $1.38 \Msun$ of
which $0.63 \Msun$ is incinerated to \Ni; the total kinetic energy is
$1.3 \times 10^{51}$ ergs. The initial model comprised 172 zones and
represented conditions $\sim 3$ s after the explosion. To reduce
subsequent computational effort, the model was evolved to $t \sim 1$
hr (using gray radiative transfer with everywhere $\kappa = 0.02
\mbox{ cm}^2\mbox{ g}^{-1}$) and it was then rezoned down to 40 zones.
This rezoning, which conserved total thermal and kinetic energy, was
nonuniform: in particular, the outermost three zones were left unaltered
in order to maintain resolution out to a maximum velocity of
$2.3\times 10^9 \mbox{ cm}\mbox{ s}^{-1}$. The density and velocity
profiles of the model at $t \sim 1$ hour are shown in Figure 1. Also
indicated is the distribution of zones. Even with only 40 zones, it
was felt that further compromise with respect to composition was
necessary. Accordingly, eight representative compositions were used, and
each of the 40 zones was assigned to one of these compositions. In the
actual computations, some interpolation between compositions was
performed, using $<Z/A>$. Though arbitrary, this interpolant proved
adequate. The compositions used for our 40 zone model are given in
Table 1.

Two benchmark calculations were done for our representation of W7 with
a hydrodynamics code that implemented radiative transfer in the
flux-limited diffusion approximation. This code is essentially the
same as that used by Sutherland \& Wheeler (1984). The first
calculation employed a constant (throughout the ejecta, and in time)
opacity of $\kappa = 0.1
\mbox { cm}^2\mbox{ g}^{-1}$. This value is typical for ``successful''
carbon deflagration models of SNIa's as calculated and judged in the
early 1980's. The second model used the tables of Rosseland means for
the LANL continuum opacities. The essential results of these
benchmarks are given in the first two lines of Table 2 and in Figure
2. The B flux and color given in the table (and plotted as B' in
Figure 2) are obtained as in Sutherland \& Wheeler (1984) by the
following prescription: (1) the photosphere is found by integrating
inward until $\tau = 2/3$ is reached; (2) the flux there, $L/4\pi
R_{ph}^2$, is equated with that of a blackbody spectrum truncated
shortward of $\lambda = 400$ nm (the temperature of this spectrum must
be determined iteratively because of the truncation), and (3) the B
and V fluxes are computed for this emitting area and temperature. This
prescription increases the B flux by $\sim 1$ mag. However,
without this or a similar rule, there would be no way to mimic the
observed strong uv deficiency and the calculated colors would always
be significantly too red. Part of the motivation for the {\em new}
code described in this paper is that light curves calculated with gray
opacities were either inadequate or uncertainly dependent upon
somewhat arbitrary prescriptions for computing B and V. The second
benchmark calculation, with the Rosseland means of the LANL opacities,
shows that $\kappa \sim 0.03 \mbox { cm}^2\mbox{ g}^{-1}$ at the
photosphere. The rise to bolometric maximum is particularly rapid (8.2
days) and reflects this lower opacity.

In the following model calculations done with the new code, the
wavelength (frequency) range used for the radiation was 7--2000 nm
($1.5\times 10^{14} - 4.3\times 10^{16}$ Hz). The colors associated
with the emergent fluxes (in the frame of the observer) were computed
with the filters given by Bessel (1990) with zero-points determined by
the computed spectrum of Vega given by Dreiling \& Bell (1980). The
colors computed in this manner never differed by more than 0.1
magnitude from those computed with single-point filters (Allen 1973,
p. 202).

Four variants were calculated for W7, reflecting different assumptions
about or treatments of the frequency dependence of the opacity. The
results are given in Figures 3-8 and summarized in Table 2. The first
variant, the ``full opacity'' model, uses all the information
available for the absorptive and scattering opacities. For the second
and third models, a simple form for the frequency dependence of
$\kappa_\nu$ was adopted:
\beq
\kappa_\nu = \kappa_0 [\theta (\lambda - \lambda_c) + 30
\theta(\lambda_c  - \lambda)]
\eeq
and the critical wavelength was taken to be either $\lambda_c = 100$
nm or $\lambda_c = 400$ nm. The enhancement factor of 30 for $\lambda
< \lambda_c$ is arbitrary and could have been much higher (see, for
example, Fig. 3 and the curve of the ``real'' $\kappa_\nu$ at the
photosphere at maximum light). The coefficient $\kappa_0$ is set by
requiring that the flux-weighted mean of the model $\kappa_\nu$ is the
same as the tabulated Rosseland mean for the real opacity. (Whenever
flux- or energy-weighted means are required in the code, they are
obtained from the frequency-dependent radiation moments as evaluated
at the previous time step.) The $\lambda_c = 100$ nm choice is meant
to simulate the very significant increase in $\kappa_\nu$ due to
bound-free transitions at short wavelengths. The $\lambda_c=400$ nm
choice is an attempt to model the effect upon the UV of the line
opacity (line blocking and expansion). However, in the absence of a
detailed calculation of the enhancement due to this opacity,
$\kappa_0$ was still set by requiring the flux-weighted opacity to
match the known, static, Rosseland mean. As a consequence of the step
up in $\kappa_\nu$ for $\lambda<\lambda_c = 400$ nm (where much of the
radiation flux is to be found), the opacity at longer wavelengths is
suppressed relative to that for the $\lambda_c=100$ nm or full opacity
models. The fourth and final model used a gray opacity. That is, the
code was run as before (multigroup radiative dynamics) but with a
constant $\kappa_\nu = \kappa_R$, the Rosseland mean. This calculation
is then very similar in spirit to those discussed by Ensman \& Burrows
(1992), although here the radiation moments are obtained at all
frequencies/colors.

Table 2 gives the essential results for the properties of the light
curves. All four models show essentially the same rise time to maximum
light ($\sim 11.5$ days for the bolometric curves with additional
delays $\lae 1$ day in B and V). This is consistent with what is known
for rise times of observed SNIa's, with the exception of the
slow-rising SN 1990N which differs in other, spectroscopic, ways
(Leibundgut et al. 1991a) from standard SNIa's such as SN 1972E and SN
1981B, and may not be interpretable with a model for the explosion
like W7. However, there is a growing consensus in the SN community
that the rise times for canonical SNIa's may be $\gae 15$ days.  If
this is true, then W7 may not be an acceptable model for SNIa's except
insofar as its velocity and composition profiles are appropriate for
spectra near maximum light.

All four models have nearly the same $M_{bol}$ at maximum light, but
this is consistent with them having identical energy deposition rates,
very nearly the same integrated opacities, and consequently similar
rise times. Figures 3-6 show, however, that the curves for $S_{rad}$
(the total instantaneous rate of energy deposition from radioactivity)
do {\em not} pass precisely through the bolometric peaks. The main
differences among the models are to be found in their colors at
maximum light and the rates of decline of these colors after maximum
light. Two of the best-studied, and by presumption most
representative, SNIa's are SN 1972E and SN 1981B for which the
following obtain (see Leibundgut et al. 1991b): at maximum light (B-V)
= -0.08 and -0.03 (respectively), (U-B) = -0.34 and -0.14, and in both
cases the rate of decline of the B curve is 2 mag in about 23
days. The full opacity model is the faintest in B, by 0.5 mag,
has too much U flux at maximum light, and is too slow to decline in B.
This suggests that if the underlying model of the explosion is
reasonable, then an enhancement of the opacity in the UV is essential.
This presumably is the effect of line blanketing and line blocking.
The properties of the $\lambda_c=100$ nm model are similar to those of
the full opacity model, except for the more rapid (and thus desirable)
decline in B. The functional {\em form} of $\kappa_\nu$ in this model
is realistic, but the neglect of scattering for $\lambda>\lambda_c$
leads to a thermalization of the radiation at lesser depths in the
SN atmosphere which in turn leads to a lower color temperature
and the more rapid decline of B. The $\lambda_c=400$ nm model is too
extreme in that U is overly suppressed and B-V is too large at maximum
light. The B maximum is the largest of the 4 models, but again this is
because the assumed form of $\kappa_\nu$ has pushed the flux into the
B and V bands. The results for the gray opacity model are virtually
identical to those of the $\lambda_c=100$ nm model. This is because,
at and after maximum light, the radiation near the photosphere (always
defined by $\tau_F=2/3$ where $\tau_F$ is the optical depth measured
inward for the flux-weighted opacity) has predominantly
$\lambda>\lambda_c$ and thus the two opacities are operationally
nearly the same.

The conclusion to be reached from these model calculations is that if
something like W7 is a good representation of the explosion, an
enhancement of the {\em scattering} opacity in the uv is probably
essential to explain the observations. This may well be the expansion
opacity, as has been argued by H\"{o}flich, M\"{u}ller, \& Khoklov
(1993), coupled with a treatment of line blanketing.  A further
conclusion is that light curve calculations for model W7 using fully
self-consistent, frequency-dependent radiation dynamics yields values
of $M_{B,max}$ that are fainter by $\sim 0.5$ magnitudes than earlier
work (Sutherland \& Wheeler 1984) suggested, although in such earlier
calculations the conversion from the bolometric light curve to the B
light curve was {\em ad hoc} and somewhat suspect. The new results are
more in keeping with larger, rather than lower, values of $H_0$ as
suggested by the recalibration of the distance to IC 4182 and the
brightness of SN 1937C (Pierce, Ressler, \& Shure 1992).

\acknowledgements{
We wish to thank Bob Clark of Los Alamos National Laboratory who
prepared the extensive opacity tables used in our calculations.  Bruno
Leibundgut provided us with light curve templates for SNIa's and
commented usefully on the uncertainties associated with UBV colors for
SNs. Robert Harkness, Doug Swartz, and Craig Wheeler gave us a
set of Rosseland mean opacities that we used in preliminary
calculations, and they also had valuable insights regarding the expansion
opacity (although this matter remains to some degree confused). The
final iteration of this paper was completed at the Aspen Center for
Physics, and any improvements can be attributed in part to the
workshop on SN spectra held there in 1993 June. This research
was supported by a grant from the Natural Sciences and Engineering
Research Council of Canada.}

\clearpage
\begin{table*}
\small
\begin{center}
\begin{tabular}{lllllllll}
\multicolumn{9}{c}{Compositions used for W7 Model}\\
    &   1 &   2 &   3
    &   4 &   5 &   6
    &   7 &   8  \\
\tableline
$^4\mbox{He}$&0.211E-08&0.162E-04&0.145E-01&0.158E-07&
     0.100E-09&0.100E-09&0.100E-09&0.100E-09\\
$^{12}\mbox{C}$& 0.100E-09&0.100E-09&0.805E-05&0.769E-08
     &0.122E-05&0.977E-05&0.100E-04&0.807E-03\\
$^{16}\mbox{O}$& 0.100E-09&0.100E-09&0.844E-08&0.466E-07
     &0.154E-04&0.664E-01&0.579&    0.582    \\
$^{20}\mbox{Ne}$&0.100E-09&0.100E-09&0.159E-07&0.100E-09
     &0.670E-09&0.263E-05&0.120E-04&0.290E-02\\
$^{24}\mbox{Mg}$&0.100E-09&0.100E-09&0.383E-07&0.562E-07
     &0.124E-04&0.142E-03&0.119&    0.122    \\
$^{26}\mbox{}$Al&0.100E-09&0.100E-09&0.353E-07&0.114E-09
     &0.104E-08&0.255E-05&0.226E-02&0.425E-02\\
$^{28}\mbox{Si}$&0.666E-08&0.100E-09&0.780E-06&0.876E-02
     &0.499&    0.561&    0.222&    0.210    \\
$^{30}\mbox{P}$& 0.198E-09&0.100E-09&0.480E-07&0.282E-06
     &0.283E-05&0.674E-03&0.535E-03&0.676E-03\\
$^{32}\mbox{S}$&0.533E-07&0.100E-09&0.223E-05&0.143E-01
    &0.298&    0.276&    0.636E-01&0.632E-01\\
$^{34}\mbox{Cl}$&0.773E-09&0.100E-09&0.152E-05&0.118E-06
    &0.215E-05&0.797E-03&0.120E-03&0.662E-03\\
$^{36}\mbox{Ar}$&0.165E-06&0.202E-09&0.565E-05&0.717E-02
    &0.556E-01&0.526E-01&0.102E-01&0.106E-01\\
$^{38}\mbox{K}$ &0.287E-08&0.100E-09&0.250E-05&0.499E-06
    &0.920E-05&0.102E-02&0.156E-04&0.131E-03\\
$^{40}\mbox{Ca}$&0.271E-07&0.189E-07&0.288E-04&0.156E-01
    &0.439E-01&0.226E-01&0.677E-03&0.650E-03\\
$^{44}\mbox{Ti}$&0.642E-05&0.181E-08&0.518E-04&0.220E-04
    &0.263E-04&0.239E-03&0.317E-05&0.288E-05\\
$^{48}\mbox{Cr}$&0.515E-01&0.144E-05&0.383E-04&0.971E-03
    &0.935E-03&0.945E-03&0.579E-05&0.548E-05\\
$^{50}\mbox{Mn}$&0.339E-02&0.802E-06&0.533E-07&0.710E-04
    &0.641E-04&0.157E-03&0.332E-06&0.144E-05\\
$^{52}\mbox{Fe}$&0.799&    0.159E-01&0.691E-04&0.719E-01
    &0.638E-01&0.171E-01&0.503E-03&0.121E-02\\
$^{54}\mbox{Co}$&0.146    &0.200&    0.815E-01&0.323E-01
    &0.832E-02&0.106E-02&0.111E-02&0.402E-03\\
$^{56}\mbox{Ni}$&0.892E-07&0.784&    0.876&    0.849
    &0.299E-01&0.817E-06&0.934E-09&0.000E+00\\
\end{tabular}
\end{center}

\caption{Mass fractions of significant elements in the eight different
compositions used in our representation of W7. In the 40 zone model,
the actual initial mass fraction of \Ni\ was used, rather than
interpolation among the eight values above.}

\end{table*}
\clearpage

\normalsize
\begin{table*}
\begin{center}
\begin{tabular}{llllllll}
Model & $\mbox{t}_{bol,max}$ & $\mbox{M}_{bol}$ & $\mbox{t}_{B,max}$ &
$\mbox{B}_{max}$ & B-V & U-B & $\delta t_2 $\\
\tableline
 FLD: $\kappa=0.1 \mbox{ cm}^2\mbox{ g}^{-1}$ & 15.1 & -19.1 & 18.8 &
-19.4 & +0.17 & N/A & $>35$ \\
 FLD: Rosseland Tables & 8.22 & -19.47 & 15.1 & -19.24 & +0.07 & N/A & 36.3\\
 Full opacity        & 11.2  & -19.7 & 12.2 & -18.9 & 0.03 & -0.87 & 32.6\\
 $\lambda_c =100$ nm & 11.5  & -19.6 & 12.2 & -18.9 & 0.14 & -0.77 & 24.7\\
 $\lambda_c =400$ nm & 11.5  & -19.6 & 11.5 & -19.4 & 0.23 & +0.15 & 19.8\\
 Gray                & 11.5  & -19.6 & 12.2 & -19.0 & 0.14 & -0.77 & 25.0\\
\tableline
\end{tabular}
\end{center}

\caption{Summary of Light Curve Calculations.
The first column identifies the opacity model. The second and third
columns give the time at, and magnitude for, bolometric maximum light.
The fourth and fifth columns give the time and maximum for the B light
curve, followed by the colors B-V and U-B at this time. The last
column is the additional time beyond that of B maximum light for the B
light curve to drop by 2 mag. The first two models were calculated
using the flux-limited diffusion (FLD) approximation as implemented by
Sutherland \& Wheeler (1984); for the first one the opacity is
everywhere constant at $\kappa = 0.1 \mbox { cm}^2\mbox{ g}^{-1}$,
while for the second the tabulated Rosseland mean opacities were
employed. For these two models the colors were computed based upon an
assumed truncated (shortward of 400 nm) blackbody spectrum, so that
the U flux is not relevant.}

\end{table*}

\clearpage


\begin{figure}
\caption {Density and velocity profiles for  model W7 of
Nomoto, Thielemann, \& Yokoi (1984). Also shown (crosses) are
the location of the 40 zone boundaries. Note how the inner and outer
zones have been set to preserve the velocity resolution of the
original 172 zone model.}
\end{figure}

\begin{figure}
\caption { Light curves for the two models calculated with
the flux-limited diffusion approximation. For the first model (upper
panels) the opacity is $\kappa = 0.1 \mbox { cm}^2\mbox{ g}^{-1}$
everywhere whereas for the second model the Rosseland mean opacities
are employed. The B light curve is based upon a blackbody spectrum
appropriate to the photospheric radius and total luminosity whereas
the B' light curve is based upon an assumed truncated (shortward of
400 nm) blackbody spectrum with the same net flux.}
\end{figure}

\begin{figure}
\caption {Light and color curves and photospheric properties
for the ``full opacity'' model. The upper left panel gives the UBV and
bolometric light curves and the total instantaneous rate of deposition
of radioactive energy ($S_{rad}$). The lower left panel gives the
radius (in $10^{15}$ cm) of the photosphere and its temperature (in
$10^4$ K) and the zone ($\triangle$) containing the photosphere. The
lower right panel gives the opacity (absorption and total) as a
function of wavelength at maximum light at the photosphere. That
$\kappa_{abs}$ slightly exceeds $\kappa_{total}$ near 600 A reflects
the limited resolution of our interpolation table.}
\end{figure}

\begin{figure}
\caption {Same as Figure 3, except for the $\lambda_c = 100$ nm model.}
\end{figure}

\begin{figure}
\caption {Same as Figure 3, except for the $\lambda_c = 400$ nm model.}
\end{figure}

\begin{figure}
\caption {Same as Figure 3, except for gray model.}
\end{figure}

\begin{figure}
\caption {Continuum spectra for the four models at bolometric
maximum light. In each panel a dashed curve for a blackbody spectrum
at the matter temperature at the photosphere is also given. The units
of $L_\nu$ are $\mbox{ergs} \mbox{ cm}^{-2} \mbox{ s}^{-1} \mbox{
Hz}^{-1}$.}
\end{figure}

\begin{figure}
\caption {A direct comparison of the four continuum spectra at
bolometric maximum light.}
\end{figure}



\clearpage
\begin{references}

\reference Allen, C.W. 1973, Astrophysical Quantities 3rd ed.
(London: Athlone)

\reference Ambwani, K., \& Sutherland, P.G. 1988, ApJ, 325, 820

\reference Arnett, W.D. 1969, Ap\&SS, 5, 180

\reference Arnett, W.D., Branch, D., \& Wheeler, J.C. 1985, Nature, 314, 337

\reference Bessel, M.S. 1990, PASP, 102, 1181

\reference Branch, D. 1990, in Supernovae, ed. A. Petschek (New York:
Springer-Verlag), 30

\reference Branch, D., Buta, R., Falk, S.W., McCall, M.L., Sutherland,
P.G., Uomoto, A., Wheeler, J.C., \& Wills, B.J. 1982, ApJ, 252, L61

\reference Branch, D., Doggett, J.B., Nomoto, K., \& Thielemann, F.-K.
1985, ApJ, 294, 619

\reference Dreiling, L.A., \& Bell, R.A. 1980, ApJ, 241, 736

\reference Ensman, L. 1991, Ph.D. thesis, Univ. of California, Santa Cruz

\reference Ensman, L., \& Burrows, A. 1992, ApJ, 393, 742

\reference Falk, S.W., Jr., \& Arnett, W.D. 1977, ApJS, 33, 515

\reference Harkness, R.P. 1986, in Radiation Hydrodynamics in Stars \&
Compact Objects, ed. D. Mihalas \& K-H A Winkler (Berlin: Springer), 166

\reference Harkness, R.P. 1991, in SN 1987A and Other Supernovae, ed.
J.~Danziger \& K.~Kj\"{a}r (Garching: ESO), 447

\reference H\"{o}flich, P., Khokhlov, A., \& M\"{u}ller, E. 1991,
A\&A, 248, L7

\reference H\"{o}flich, P., M\"{u}ller, E., \& Khokhlov, A. 1993,
A\&A, 268, 570

\reference Huebner, W.F., Merts, A.L., Magee, N.H.,\& Argo, M.F.
1977, Los Alamos Sci. Lab. Report LA-6760-M

\reference Jacoby, G.H., et al. 1992, PASP, 104, 599

\reference Karp, A.H., Lasher, G., Chan, K.L., \& Salpeter, E.E.
1977, ApJ, 214, 161

\reference Khokhlov, A. 1991, A\& A, 245, 114

\reference Khokhlov, A., M\"{u}ller, E. \& H\"{o}flich, P. 1992,
A\&A, 253, L9

\reference Khokhlov, A., M\"{u}ller, E. \& H\"{o}flich, P. 1993,
A\&A, 270, 223

\reference Leibundgut, B., Kirshner, R.P., Filippenko, A.V.,
Shields, J.C., Foltz, C.B., Phillips, M.M., \& Sonneborn, G.
1991a, ApJ, 371, L23

\reference Leibundgut, B., \& Tammann, G.A. 1990, A\& A, 230, 81

\reference Leibundgut, B., Tammann, G.A., Cadonau, R., \& Cerrito, D.
1991b, A\&AS, 89, 537

\reference Mihalas, D., \& Mihalas, B.W. 1984, Foundations
of Radiation Hydrodynamics (Oxford: Oxford Univ. Press)

\reference Magee, N.H., Merts, A.L., \& Huebner, W.F. 1984, ApJ, 283, 264

\reference Nomoto, K., \& Shigeyama, T. 1991, in Supernovae, 10th
Santa Cruz Summer Workshop in Astronomy and Astrophysics, ed. S.
Woosley (New York: Springer-Verlag), 572.

\reference Nomoto, K., Sugimoto, D., \& Neo, S. 1976, Ap\&SS, 39, L37

\reference Nomoto, K., Thielemann, F.-K., \& Yokoi, K. 1984, ApJ, 286, 644

\reference Pierce, M.J., Ressler, M.E., \& Shure, M.S. 1992, ApJ, 390, L45

\reference Rogers, F.J., \& Iglesias, C.A. 1992, ApJS, 79, 507


\reference Sutherland, P.G., \& Wheeler, J.C. 1984, ApJ, 280, 282

\reference Swartz, D.A., Harkness, R.P. \& Wheeler, J.C. 1991, ApJ, 374, 266

\reference Wheeler, J.C., \& Hansen, C.J. 1971, Ap\& SS, 11, 373

\reference Wheeler, J.C., \& Harkness, R.P. 1990, Rep. Prog. Phys., 53, 1467

\reference Woosley, S.E. 1990, in Supernovae, ed. A. Petschek (New
York: Springer Verlag), 182

\end{references}
\end{document}